\documentclass[aps,twocolumn,pra,superscriptaddress,tightenlines]{revtex4}
\usepackage{epsfig,graphicx,times}
\usepackage{amstext}
\usepackage{amsmath}
\usepackage{amssymb}
\usepackage{graphicx}
\usepackage{latexsym}
\usepackage{bm}
\usepackage[colorlinks,citecolor=blue, linkcolor=blue,hyperindex,CJKbookmarks,dvipdfm]{hyperref}

\begin{document}

\title{Cross-correlation between photons and phonons in quadratically
coupled optomechanical systems}
\author{Xun-Wei Xu}
\email{davidxu0816@163.com}
\affiliation{Department of Applied Physics, East China Jiaotong University, Nanchang,
330013, China}
\author{Hai-Quan Shi}
\affiliation{School of Materials Science and Engineering, Nanchang University, Nanchang
330031, China}
\affiliation{Department of Applied Physics, East China Jiaotong University, Nanchang,
330013, China}
\author{Ai-Xi Chen}
\email{aixichen@ecjtu.edu.cn}
\affiliation{Department of Physics, Zhejiang Sci-Tech University, Hangzhou, 310018, China}
\affiliation{Department of Applied Physics, East China Jiaotong University, Nanchang,
330013, China}
\author{Yu-xi Liu}
\affiliation{Institute of Microelectronics, Tsinghua University, Beijing 100084, China}
\affiliation{Beijing National Research Center for Information Science and Technology (BNRist), Beijing 100084, China}
\date{\today }

\begin{abstract}
We study photon, phonon statistics and the cross-correlation
between photons and phonons in a quadratically coupled optomechanical
system. Photon blockade, phonon blockade, and strong anticorrelation between photons and phonons can be observed in the same parameter regime with the effective
nonlinear coupling between the optical and mechanical modes, enhanced by a
strong optical driving field. Interestingly, an optimal value of the effective
nonlinear coupling strength for the photon blockade is not within the strong
nonlinear coupling regime. This abnormal phenomenon
results from the destructive interference between different paths for
two-photon excitation in the optical mode with a moderate effective
nonlinear coupling strength. Furthermore, we show that phonon (photon)
pairs or correlated photons and phonons can be generated in the strong
nonlinear coupling regime with a proper detuning between the weak mechanical driving field and mechanical mode. Our results open up a way to generate anticorrelated and correlated photons and phonons, which may have
important applications in quantum information processing.
\end{abstract}

\maketitle



\section{Introduction}

To fulfill the needs of quantum information processing~\cite%
{ScaraniRMP09,JWPanRMP12}, the development of single photon sources~\cite%
{LounisRPP05,BuckleyRPP12,GCShanFP14} has become a research focus in quantum
physics. A perfect single photon source can emit a single photon in one time
so that the emitted photons show strong antibunching effect. Photon blockade~%
\cite{ImamogluPRL97} that the excitation of the first photon blocks the
excitation of the second photon can be used to realize
perfect single photon source. Photon blockade has been observed in many
different systems, such as cavity quantum electrodynamics systems~\cite%
{BirnbaumNat05,DayanSci08,DubinNP10}, a quantum dot in a photonic crystal~%
\cite{FaraonNPy08,XDingPRL16}, and circuit quantum electrodynamics system~\cite%
{CLangPRL11,HoffmanPRL11,LiuPRA14}.

As the counterpart of photons, phonons are the elementary excitations in
mechanical systems. With the development of micro- and nano-technology, such
as improving the quality factor of the mechanical oscillators and enhancing
the coupling strengths of mechanical oscillators to other quantum systems~%
\cite{SchwabPT05,PootPR12}, the appearance of phonon lasers~\cite%
{VahalaNPy09,GrudininPRL10,HWangPRA14} and theoretical proposals on
generating single phonons~\cite%
{YXLiuPRA10,DidierPRB11,RamosPRL13,CohenNature15,MiranowiczPRA16,XWXuPRA16,XWangARX16,BarzanjehPRA16,XWXuPRA13a}%
, phonons gradually become a new candidate for quantum information
processing. Moreover, the measurement of the correlations of the phonons can
be realized by converting the mechanical signals into optical signals
through auxiliary optomechanical couplings~\cite%
{RamosPRL13,CohenNature15,XWXuPRA16}. In a recent experiment~\cite%
{CohenNature15}, the phonon correlation in
an optomechanical system has been measured by detecting the correlations of
the emitted photons from the optical cavity.

As the complexity of the tasks, entrusted to the quantum information
processors, becomes higher and higher, hybrid optical and mechanical systems
become more and more important ~\cite%
{WallquistPS09,XiangZLRMP13}. One useful approach to
improve the overall coordination is to design a hybrid device which can
generate not only single photons and single phonons, but also correlated
single photons and single phonons. The coupling between optical and
mechanical systems based on two main ways: one is induced indirectly by
muti-level atoms, such as artificial atoms based superconducting quantum
circuits~\cite{XiangZLRMP13,XGuPR17} and nitrogen-vacancy centers~\cite%
{ArcizetNPy11,MacQuarriePRL13}; the other one is induced by parameters
coupling, such as optomechanical coupling~\cite{CKLawPRA94}.

Optomechanical system, that a cavity mode is coupled to a mechanical mode
via radiation pressure or optical gradient forces, provides us an
appropriate platform to manipulate both photons and
phonons simultaneously (for reviews, see Refs.~\cite%
{KippenbergSci08,MarquardtPhy09,AspelmeyerPT12,AspelmeyerARX13,MetcalfeAPR14,YLLiuCPB18}%
). It has been shown that photon blockade can be realized in the
optomechanical system with different structures, such as
strongly coupled optomechanical systems at single-photon level~\cite%
{RablPRL11,NunnenkampPRL11,KronwaldPRA13,XWXuPRA13,JQLiaoPRA13a,JQLiaoPRA13b,DHuPRA15}%
, multimode optomechanical systems~\cite%
{StannigelPRL12,KomarPRA13,XWXuJPB13,SavonaarX13}, squeezed
optomechanical systems~\cite{XYLvPRL15,XYLvPRA17},
quadratically coupled optomechanical system driven by a strong optical field~\cite{HXiePRA16}.
Recently, strong phonon antibunching was proposed in a quadratically
coupled optomechanical system~\cite{SeokPRA17,HXiePRA17,HQShiSR18} and this provides another
possible way to generate single phonons in the optomechanical system.
However, the realization of both photon blockade and phonon blockade in one
optomechanical system with the same parameters has not been considered yet.
Moreover, optomechanical system provides us an ideal platform to investigate
the cross-correlation between photons and phonons~\cite{CarligPRA14}, which has important
applications in quantum information processing.

In this paper, we study photon statistics, phonon statistics and the
cross-correlation between photons and phonons in a quadratically coupled
optomechanical system. The effective nonlinear coupling between the optical and mechanical modes in a quadratically coupled optomechanical system can be enhanced by a strong driving optical field as shown in Refs.~\cite{HXiePRA16,HXiePRA17,HQShiSR18,LGSiPRA17}.
Different from the previous studies~\cite{HXiePRA16,HXiePRA17}, we find that there is an optimal value of the effective nonlinear coupling strength for photon blockade before reaching the strong
nonlinear coupling regime, and both photon and phonon blockades can be observed in a quadratically coupled
optomechanical system with the same parameters.
Moreover, we also study the cross-correlation between photons and phonons, and show that both anticorrelated and correlated phonons and photons can be generated in the quadratically coupled optomechanical system.

The paper is organized as follows. In Sec.~II, we
show the theoretical model of the quadratically coupled optomechanical
system. In Sec.~III, we study phonon blockade, photon blockade,
anti-correlation between phonons and photons, and the effects of the
parameters on the statistical properties of the system are discussed. In
Sec.~IV, we explain how the strong photon blockade appears with a weak
nonlinear coupling strength. In Sec.~V, we show that phonon (photon) pairs
and correlated photons and phonons can be generated in the strong nonlinear
coupling regime with a proper detuning between the weak mechanical driving field and mechanical mode. Finally, we summarize the results in
Sec.~VI.

\section{Theoretical model}

We study an optomechanical system in which a mechanical mode is quadratically
coupled to an optical mode. Such system can be found in the optomechanical
crystals~\cite{ParaisoPRX15}, Fabry-Perot cavities with
membrane-in-the-middle~\cite%
{ThompsonNat08,Flowers-JacobsAPL12,HXuNat16,HXuNC17}, and other
optomechanical systems~\cite{PurdyPRL10,JTHill13,DoolinPRA14,BrawleyNC16}.
We assume that the optical mode
is driven by an external field with the strength $\Omega $ and frequency $%
\omega _{L}$, and the mechanical mode is driven by a mechanical pump of
strength $2\sqrt{2}\varepsilon \cos \left( \omega _{d}t\right) $ (amplitude $%
\varepsilon $, frequency $\omega _{d}$).
In the rotating reference frame with optical frequency $\omega _{L}$, the
system can be described by a Hamiltonian ($\hbar =1$)%
\begin{eqnarray}
H &=& \Delta _{c} A^{\dag }A+\frac{1}{2}\omega
_{m}\left( Q^{2}+P^{2}\right) +2gA^{\dag }AQ^{2}  \notag \\
&&+\left( \Omega A^{\dag }+\mathrm{H.c.}\right) +2\sqrt{2}Q\varepsilon \cos
\left( \omega _{d}t\right),   \label{Eq1}
\end{eqnarray}%
where $\Delta _{c}=\omega _{c}-\omega _{L}$ is the detuning of the strong optical driving field from the optical mode with frequency $\omega _{c}$; $A$ and $A^{\dag }$ are the annihilation and creation operators of the
optical mode, $Q$ and $P$ are the
dimensionless displacement and momentum operators of the mechanical mode
with frequency $\omega _{m}$, and $g>0$ is the quadratic optomechanical
coupling strength between the optical and mechanical modes. The
damping rates of the optical mode and mechanical mode are $\gamma _{c}$ and $%
\gamma _{m}$, respectively. We assume that the strength of the optical driving field is
strong, i.e., $\Omega \gg \gamma _{c}$, while the strength of the mechanical
driving field is much weaker than the damping rate of the optical mode,
i.e., $\varepsilon \ll \gamma _{c}$.

\begin{figure}[tbp]
\includegraphics[bb=27 265 561 724, width=8.5 cm, clip]{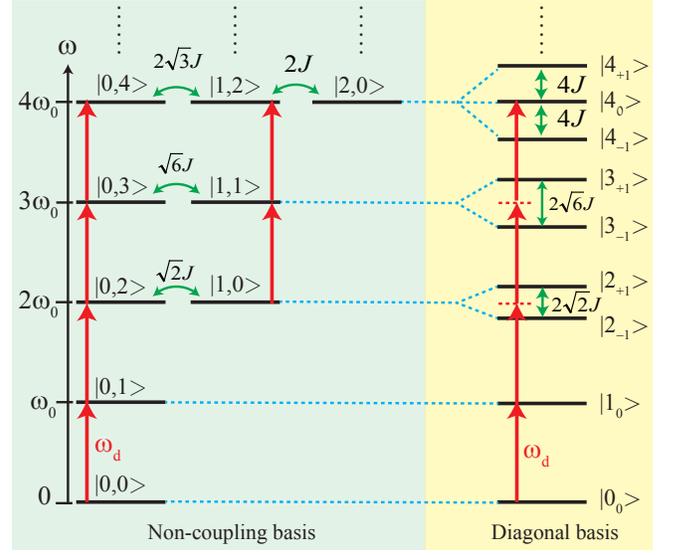}
\caption{(Color online) The schematic energy spectrum of the linearized
quadratically coupled optomechanical system [see the Hamiltonian in Eq.~(%
\protect\ref{Eq5})] given in the non-coupling basis (left) and in the
diagonal basis (right).}
\label{fig1}
\end{figure}

Based on the Hamiltonian in Eq.~(\ref{Eq1}), a tunable second-order nonlinear coupling between the optical and mechanical modes can be induced by the strong optical driving field. The quantum Langevin equations (QLEs) for the operators are given by%
\begin{equation}
\frac{dA}{dt}=-\left( i\Delta _{c}+\frac{\gamma _{c}}{2}\right)
A-i2gAQ^{2}-i\Omega +\sqrt{\gamma _{c}}A_{\rm in},
\end{equation}%
\begin{equation}
\frac{dQ}{dt}=\omega _{m}P,
\end{equation}%
\begin{equation}
\frac{dP}{dt}=-\omega _{m}Q-4gA^{\dag }AQ-\frac{\gamma _{m}}{2}P-2\sqrt{2}%
\varepsilon \cos \left( \omega _{d}t\right) +\xi,
\end{equation}%
where $A_{\rm in}$ and $\xi$ are the noise operators with zero mean values. The steady state mean values $\alpha $, $Q_{s}$ and $%
P_{s}$ of operators $A$, $Q$ and $P$ can be obtained by taking the quantum average of the QLEs and setting the time derivatives to zeros. Without considering the weak mechanical driving field, i.e., $%
\varepsilon =0$, the steady state mean values $\alpha $, $Q_{s}$ and $P_{s}$ for $g>0$ are shown as
\begin{equation}
\alpha =\frac{-i2\Omega }{\gamma _{c}+i2\Delta _{c}},  \label{Eq2}
\end{equation}%
\begin{equation}
Q_{s}=P_{s}=0.  \label{Eq3}
\end{equation}%
We expand the operators as the sum of their steady state mean values and quantum
fluctuations: $A\rightarrow \alpha +a$, $Q\rightarrow Q_{s}+q$ and $%
P\rightarrow P_{s}+p$, where $a$, $q$ and $p$ are the quantum
flucturation operators, then the effective Hamiltonian $H^{\prime }$ for the quantum
flucturation operators reads
\begin{eqnarray}
H^{\prime } &=&\Delta _{c}a^{\dag }a+\omega _{m}b^{\dag }b+g\left(
\left\vert \alpha \right\vert ^{2}+a^{\dag }a\right) \left( b^{\dag
}+b\right) ^{2}  \notag \\
&&+g\left( \alpha a^{\dag }+\alpha ^{\ast }a\right) \left( b^{\dag
}+b\right) ^{2}  \notag \\
&&+2\varepsilon \cos \left( \omega _{d}t\right) \left( b^{\dag }+b\right),
\label{Eq4}
\end{eqnarray}%
where $q\equiv \left( b^{\dag }+b\right) /\sqrt{2}$, $p\equiv i\left( b^{\dag }-b\right)
/\sqrt{2}$. For a strong optical driving field, we assume that the
steady-state mean value $\alpha $ is much larger than the quantum
flucturation operators $a$, such as $\left\vert \alpha \right\vert ^{2}\gg
\left\langle a^{\dag }a\right\rangle $, then the term $ga^{\dag }a\left(
b^{\dag }+b\right) ^{2}$ in the above equation can be neglected. In the
rotating reference frame with respect to the unitary operator $R\left(
t\right) =\mathrm{exp}\left( i2\omega _{d}a^{\dag }at+i\omega _{d}b^{\dag
}bt\right) $, under the rotating-wave approximation by neglecting the terms
oscillating with high frequencies in Eq.~(\ref{Eq4}), e.g. $2\omega _{d}$
and $4\omega _{d}$, a simplified effective Hamiltonian is obtained as
\begin{equation}
H_{\mathrm{eff}}=\Delta a^{\dag }a+\Delta _{m}b^{\dag }b+Ja^{\dag
}b^{2}+J^{\ast }ab^{\dag 2}+\left( \varepsilon b^{\dag }+\mathrm{H.c.}%
\right),   \label{Eq5}
\end{equation}%
where the detunings $\Delta =\Delta _{c}-2\omega _{d}$ and $\Delta
_{m}=\omega _{m}+2g\left\vert \alpha \right\vert ^{2}-\omega _{d}$ satisfy
the condition $\left\{ \left\vert \Delta \right\vert ,\left\vert \Delta
_{m}\right\vert \right\} \ll \omega _{m}$; $J=g\alpha $ is the effective
second-order nonlinear coupling strength between the optical and mechanical modes, and can be controlled by tuning the strength of the strong optical driving field.
Without loss of generality $J$ is assumed to be real.

The energy spectrum of the Hamiltonian for the linearized quadratically
coupled optomechanical system in Eq.~(\ref{Eq5}) is shown in Fig.~\ref{fig1}%
, where $\omega _{0}\equiv \omega _{m}+2g\left\vert \alpha \right\vert^{2}
=\Delta _{c}/2$. In the non-coupling basis (left), $\left\vert
n,m\right\rangle $ represents the Fock state with $n$ photons in optical
mode and $m$ phonons in the mechanical mode. In the diagonal basis (right),
we have $\left\vert 0_{0}\right\rangle \equiv \left\vert 0,0\right\rangle $,
$\left\vert 1_{0}\right\rangle \equiv \left\vert 0,1\right\rangle $, $%
\left\vert 2_{\pm 1}\right\rangle \equiv (\left\vert 1,0\right\rangle \pm
\left\vert 0,2\right\rangle)/\sqrt{2} $, $\left\vert 3_{\pm1}\right\rangle
\equiv (\left\vert 1,1\right\rangle \pm \left\vert 0,3\right\rangle)/\sqrt{2}
$, $\left\vert 4_{0}\right\rangle \equiv (-\sqrt{3} \left\vert
2,0\right\rangle + \left\vert 0,4\right\rangle)/2 $ and $\left\vert
4_{\pm1}\right\rangle \equiv (\left\vert 2,0\right\rangle \pm 2\left\vert
1,2\right\rangle + \sqrt{3} \left\vert 0,4\right\rangle)/(2\sqrt{2})$. To
have $J \sim \gamma_c$ in the weak quadratically coupling regime $g\ll
\gamma_c$, the optical mode has to be strongly driven with $\left\vert
\alpha \right\vert\gg 1$. In this case, we have the frequency shift $%
\omega^{\prime }\equiv 2g\left\vert \alpha \right\vert ^{2}\approx2\left\vert
J\alpha \right\vert \gg \gamma_c$ and the effective frequency of the
mechanical mode $\omega _{0}=\omega _{m}+\omega^{\prime }$ may be much
higher than the bare frequency of the mechanical mode $\omega _{m}$.

To quantify the statistics of the phonons and photons in the system, we
consider the second-order correlation functions in the steady state ($%
t\rightarrow \infty $) defined by
\begin{eqnarray}
g_{aa}^{\left( 2\right) }\left( \tau \right) &\equiv &\frac{\left\langle
a^{\dag }\left( t\right) a^{\dag }\left( t+\tau \right) a\left( t+\tau
\right) a\left( t\right) \right\rangle }{n_{a}\left( t\right) ^{2}}, \\
g_{bb}^{\left( 2\right) }\left( \tau \right) &\equiv &\frac{\left\langle
b^{\dag }\left( t\right) b^{\dag }\left( t+\tau \right) b\left( t+\tau
\right) b\left( t\right) \right\rangle }{n_{b}\left( t\right) ^{2}}, \\
g_{ab}^{\left( 2\right) }\left( \tau \right) &\equiv &\left\{
\begin{array}{cc}
\frac{\left\langle a^{\dag }\left( t\right) b^{\dag }\left( t+\tau \right)
b\left( t+\tau \right) a\left( t\right) \right\rangle }{n_{a}\left( t\right)
n_{b}\left( t\right) } & \tau \geq 0 \\
\frac{\left\langle b^{\dag }\left( t\right) a^{\dag }\left( t-\tau \right)
a\left( t-\tau \right) b\left( t\right) \right\rangle }{n_{a}\left( t\right)
n_{b}\left( t\right) } & \tau <0%
\end{array}%
\right. ,
\end{eqnarray}%
where $n_{a}\left( t\right) \equiv \left\langle a^{\dag }\left( t\right)
a\left( t\right) \right\rangle $ and $n_{b}\left( t\right) \equiv
\left\langle b^{\dag }\left( t\right) b\left( t\right) \right\rangle $ are
the mean photon and phonon numbers. The dynamic behavior of the total open
system is described by the master equation for the density matrix $\rho $~%
\cite{Carmichael}
\begin{eqnarray}
\frac{\partial \rho }{\partial t} &=&-i\left[ H_{\mathrm{eff}},\rho \right]
+\gamma _{c}L[a]\rho  \notag  \label{EqMaster} \\
&&+\gamma _{m}\left( n_{\mathrm{th}}+1\right) L[b]\rho +\gamma _{m}n_{%
\mathrm{th}}L[b^{\dag }]\rho
\end{eqnarray}%
where $L[o]\rho =o\rho o^{\dag }-\left( o^{\dag }o\rho +\rho o^{\dag
}o\right) /2$ denotes a Lindbland term for an operator $o$; $n_{\mathrm{th}}$
is the mean number of the thermal phonon, given by the Bose-Einstein
statistics $n_{\mathrm{th}}=[\exp (\hbar \omega _{m}/k_{B}T)-1]^{-1}$ with
the Boltzmann constant $k_{B}$ and the environmental temperature $T$. The temperature effect on the photon is neglected, because we assume that the optical frequency is much higher than the mechanical one.  The
second-order correlation functions can be calculated by solving the master
equation in Eq.~(\ref{EqMaster}) numerically within a truncated Fock space.

\section{Anticorrelated phonons and photons}

\begin{figure}[tbp]
\includegraphics[bb=39 162 549 745, width=8.5 cm, clip]{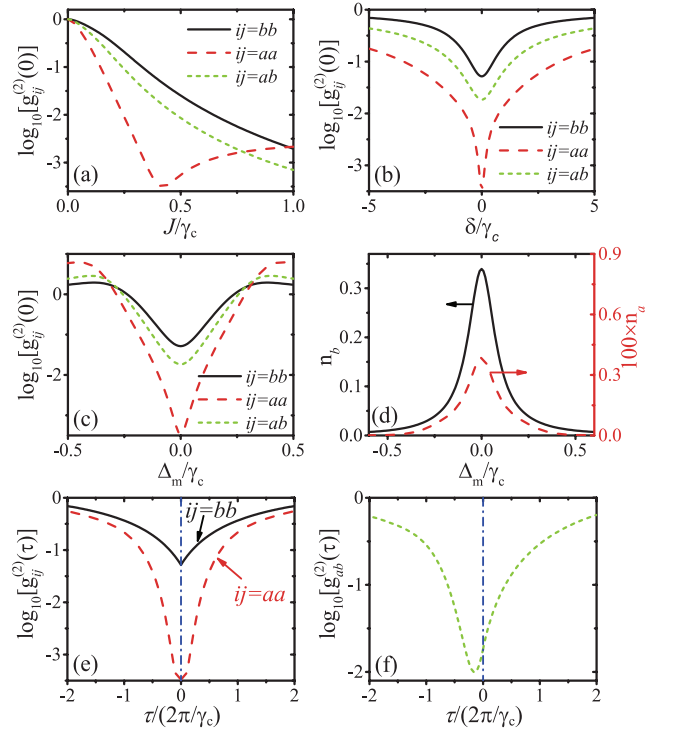}
\caption{(Color online) $\log_{10}g_{ij}^{\left(2\right) }\left( 0\right)$ ($%
ij=bb,aa,ab$) are plotted (a) as functions of the effective coupling
strength $J/\protect\gamma_c$; (b) as functions of the detuning $\protect%
\delta/\protect\gamma_c \equiv (\protect\omega_c-\protect\omega_L-2\protect%
\omega_d)/\protect\gamma_c$; (c) as functions of the detuning $\Delta_{m}/%
\protect\gamma_c$. (d) Mean phonon number $n_b$ and photon number $100\times
n_a$ are plotted as functions of the detuning $\Delta_{m}/\protect\gamma_c$. $%
\log_{10}g_{ij}^{\left( 2\right) }\left( \protect\tau\right)$ is plotted as
a function of the normalized time delay $\protect\tau/(2\protect\pi/\protect%
\gamma_c)$ in (e) and (f). $\Delta=\Delta_{m}=0$ in (a); $\protect\omega_0=%
\protect\omega_d$ and $J = 0.406\protect\gamma_c$ in (b); $J = 0.406\protect%
\gamma_c$ and $\Delta=2\Delta_{m}$ in (c) and (d); $\Delta=\Delta_{m}=0$ and
$J = 0.406\protect\gamma_c$ in (e) and (f). The other parameters are $%
\protect\varepsilon=0.05\protect\gamma_c$, $\protect\gamma_m=\protect\gamma%
_c/10$, and $n_{\mathrm{th}}=10^{-4}$.}
\label{fig2}
\end{figure}

Figure~\ref{fig2}(a) displays $\log_{10}g_{ij}^{\left(2\right) }\left(
0\right)$ ($ij=bb,aa,ab$) as functions of the effective coupling strength $%
J/\gamma_c$ for $\Delta =\Delta _{m}=0$. $\log_{10}g_{bb}^{\left(2\right)
}\left( 0\right)$ and $\log_{10}g_{ab}^{\left(2\right) }\left( 0\right)$
decrease gradually with the increase of the effective coupling strength $J$%
. Similar phenomena were mentioned in doubly resonant nanocavities with
second-order nonlinearity~\cite{MajumdarPRA13}, where a strongly
anticorrelation between the first- and second-harmonic photons was reported.
Here, we propose to realize strongly anticorrelation between photons and
phonons with quadratically optomechanical coupling, which may have important
applications in building hybrid systems. Moreover, different from the
monotone increases of $\log_{10}g_{bb}^{\left(2\right) }\left( 0\right)$
and $\log_{10}g_{ab}^{\left(2\right) }\left( 0\right)$, the minimum of $%
\log_{10}g_{aa}^{\left(2\right) }\left( 0\right)$ appears with the effective
coupling strength $J\approx 0.406\gamma_c$.

The monotone increases of $\log_{10}g_{bb}^{\left(2\right) }\left( 0\right)$
and $\log_{10}g_{ab}^{\left(2\right) }\left( 0\right)$ can be understood by
the energy spectrum shown in Fig.~\ref{fig1}. When the mechanical mode is
driven by field with the frequency $\omega_d=\omega_0$, we can realize the phonon
blockade in analogy to the cavity QED~\cite%
{BirnbaumNat05,DayanSci08,DubinNP10,FaraonNPy08,XDingPRL16,CLangPRL11,HoffmanPRL11,LiuPRA14}%
: the transition from $\left\vert 0_{0}\right\rangle$ to $\left\vert
1_{0}\right\rangle$ is enhanced with resonant phonon absorption, while the
transition from $\left\vert 1_{0}\right\rangle$ to $\left\vert
2_{\pm1}\right\rangle$ is blocked for detuning $\sqrt{2}J$. The
anti-correlation between the photons and phonons for $g_{ab}^{\left(2\right)
}\left( 0\right)<1$ can be understood in a similar way: the the transition
from $\left\vert 2_{\pm1}\right\rangle$ to $\left\vert 3_{\pm1}\right\rangle$
is also blocked for detuning $(\sqrt{6}-\sqrt{2})J$. However, the appearing
of the minimum of $\log_{10}g_{aa}^{\left(2\right) }\left( 0\right)$ with
the effective coupling strength $J\approx 0.406\gamma_c$ cannot be explained
by the same way with nonlinear energy spectrum of the system. This abnormal
phenomenon results from the destructive interference between different paths
for two-photon excitation in the optical mode~\cite{BambaPRA11} and we will
give a detailed explanation in the next section.

There are two external driving fields applied to the system simultaneously:
a strong optical field with frequency $\omega_L$ and a weak mechanical
driving field with frequency $\omega_d$. $\log_{10}g_{ij}^{\left(2\right)
}\left( 0\right)$ ($ij=bb,aa,ab$) are plotted as functions of the detuning $%
\delta/\gamma_c \equiv (\omega_c-\omega_L-2\omega_d)/\gamma_c$ in Fig.~\ref%
{fig2}(b) and detuning $\Delta_{m}=\Delta/2$ in Fig.~\ref{fig2}(c). In Fig.~%
\ref{fig2}(b), we set $\omega_0=\omega_d$ and the detunings $\delta$ is
changed by tuning the frequency $\omega_L$; in Fig.~\ref{fig2}(c), we set $%
\omega_L=\omega_c-2\omega_0$ and the detuning $\Delta_m$ ($\Delta$) is
changed by tuning the frequency $\omega_d$. The figures show that the photon
blockade, phonon blockade, and strongly anticorrelated photons and phonons
are much more rigid against the tuning of frequency $\omega_L$ but more
sensitive to the tuning of frequency $\omega_d$.

\begin{figure}[tbp]
\includegraphics[bb=60 212 558 596, width=8.5 cm, clip]{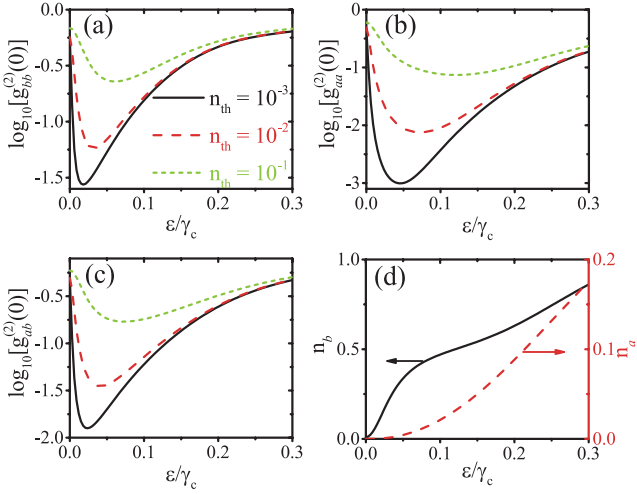}
\caption{(Color online) $\log_{10}g_{ij}^{\left( 2\right) }\left( 0\right)$
[(a) $ij=bb$, (b) $ij=aa$, (c) $ij=ab$] is plotted as a function of the
driving strength $\protect\varepsilon/\protect\gamma_c$ for different mean
thermal phonon number $n_{\mathrm{th}}$ [solid curve for $n_{\mathrm{th}%
}=10^{-3}$; dashed curve for $n_{\mathrm{th}}=10^{-2}$; dotted curve for $n_{%
\mathrm{th}}=10^{-1}$]. (d) Mean phonon number $n_b$ and photon number $n_a$
are plotted as functions of the driving strength $\protect\varepsilon/%
\protect\gamma_c$ for mean thermal phonon number $n_{\mathrm{th}}=10^{-2}$.
The other parameters are $\Delta=\Delta_{m}=0$, $J = 0.406\protect\gamma_c$,
and $\protect\gamma_m=\protect\gamma_c/10$.}
\label{fig3}
\end{figure}

Mean phonon number $n_b$ and photon number $n_a$ for photon blockade, phonon
blockade and strongly anticorrelated photons and phonons are plotted as
functions of the detuning $\Delta/\gamma_c$ in Fig.~\ref{fig2}(d). As the
generation of a single photon needs annihilating two single phonons, the
efficient for single-photon generation is much lower than the one for
single-phonon generation. The second-order correlation function $%
g_{ij}^{\left( 2\right) }\left( \tau \right)$ ($ij=bb,aa,ab$) is plotted as
a function of the normalized time delay $\tau/(2\pi/\gamma_c)$ in Fig.~\ref%
{fig2} (e) and (f) for $\Delta=\Delta_{m}=0$ and $J = 0.406\gamma_c$. The
time duration for photon blockade, phonon blockade and strongly
anticorrelated photons and phonons is of the order of the lifetime of the
photons in the cavity. The cross-correlation function for anticorrelated
photons and phonons is asymmetric for $\tau>0$ and $\tau<0$.

$\log_{10}g_{ij}^{\left( 2\right) }\left( 0\right)$ is plotted as a function
of the mechanical driving strength $\varepsilon/\gamma_c$ for different mean
thermal phonon numbers $n_{\mathrm{th}}$ in Fig.~\ref{fig3}: (a) $ij=bb$, (b)
$ij=aa$, (c) $ij=ab$. Mean phonon number $n_b$ and photon number $n_a$ are
plotted as functions of the mechanical driving strength $\varepsilon$ in
Fig.~\ref{fig3}(d). Clearly, the thermal phonons have a
detrimental effect on the realization of photon blockade, phonon blockade
and strongly anticorrelated photons and phonons. A proper increase of the
mechanical driving strength $\varepsilon$ can increase the number of mean
phonons and photons and this is also helpful to overcome the detrimental effect induced by the thermal phonons. But if the mechanical driving
strength becomes too strong, the phonons and photons tend to behave
classically.

\section{Unconventional photon blockade}

\begin{figure}[tbp]
\includegraphics[bb=20 380 549 598, width=8.5 cm, clip]{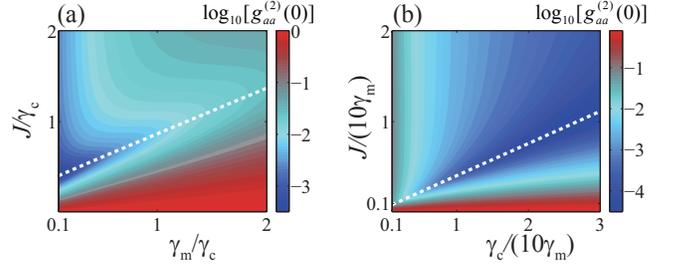}
\caption{(Color online) (a) Contour plot of $\log_{10}g_{aa}^{\left(
2\right) }\left( 0\right)$ as a function of the damping rate $\protect\gamma%
_m/\protect\gamma_c$ and the effective coupling strength $J/\protect\gamma_c$
for $\protect\varepsilon =0.05\protect\gamma_c$; (a) contour plot of $%
\log_{10}g_{aa}^{\left( 2\right) }\left( 0\right)$ vs the damping rate $%
\protect\gamma_c/(10\protect\gamma_m)$ and the effective coupling strength $%
J/(10\protect\gamma_m)$ for $\protect\varepsilon =0.5\protect\gamma_m$. The
white dashed line refers to Eq.~(\protect\ref{EqJ}). The other parameters
are $\Delta=\Delta_{m}=0$ and $n_{\mathrm{th}}=10^{-4}$.}
\label{fig4}
\end{figure}

To understand the origin of the strong photon antibunching
appearing with weak coupling strength $J\approx 0.406\gamma_c $, we now
examine the paths for two-photon excitation. As illustrated in Fig.~\ref%
{fig1} (left), there are two transition paths for two-photon generation: $%
\left\vert 0,4\right\rangle \rightarrow \left\vert 1,2\right\rangle
\rightarrow \left\vert 2,0\right\rangle$ and $\left\vert 1,1\right\rangle
\rightarrow \left\vert 1,2\right\rangle \rightarrow \left\vert
2,0\right\rangle$. The strong photon antibunching can be explained using the
destructive interference between the two different paths of two-photon
generation. The occupation probabilities in states $\left\vert
1,2\right\rangle$ and $\left\vert 2,0\right\rangle$ become zero when the
transition matrix elements of these two paths of photon excitation have the
same amplitude but different phase. To examine this explanation, following the
method given in Ref.~\cite{BambaPRA11}, we will derive the optimal
conditions for strong photon antibunching with the resonant driving
condition $\Delta =\Delta _{m}=0$.

As the mechanical driving field is not so strong that the average photon and
phonon numbers $n_{a}$ and $n_{b}$ are small, i.e. $n_{a}\ll 1$ and $n_{b}\ll
1 $, we can expand the wave function $\left\vert \psi \right\rangle $ in the Fock-state basis $\left\vert
n,m\right\rangle $ truncated to the two-photon and four-phonon states. That is, we assume
\begin{eqnarray}  \label{Eq10}
\left\vert \psi \right\rangle &=&C_{00}\left\vert 0,0\right\rangle
+C_{01}\left\vert 0,1\right\rangle  \notag \\
&&+C_{02}\left\vert 0,2\right\rangle +C_{10}\left\vert 1,0\right\rangle
\notag \\
&&+C_{03}\left\vert 0,3\right\rangle +C_{11}\left\vert 1,1\right\rangle
\notag \\
&&+C_{04}\left\vert 0,4\right\rangle +C_{12}\left\vert 1,2\right\rangle
+C_{20}\left\vert 2,0\right\rangle
\end{eqnarray}%
where the coefficients satisfy $C_{00}\approx 1\gg C_{01}\gg
C_{02},C_{10}\gg C_{03},C_{11}\gg C_{04},C_{12},C_{20}$. The
coefficient $|C_{nm}|^2$ denotes the occupying probability in the state $%
\left\vert n,m\right\rangle $. Substituting the wave function in Eq.~(\ref%
{Eq10}) and the effective Hamiltonian in Eq.~(\ref{Eq5}) with $\Delta
=\Delta _{m}=0$ into the Schr\"{o}dinger's equation $i\partial
_{t}\left\vert \psi \right\rangle =H_{\rm eff}\left\vert \psi
\right\rangle $, the dynamical equations for the coefficients $C_{nm}$ can be
obtained by taking account of the dampings of the optical and mechanical
modes. The steady-state values of the coefficients $C_{nm}$ are determined
by the equations%
\begin{equation}
0=-\frac{\gamma _{m}}{2}C_{01}-i\varepsilon C_{00},
\end{equation}%
\begin{eqnarray}
0 &=&-\frac{\gamma _{c}}{2}C_{10}-i\sqrt{2}JC_{02}, \\
0 &=&-\gamma _{m}C_{02}-i\sqrt{2}JC_{10}-i\sqrt{2}\varepsilon C_{01},
\end{eqnarray}%
\begin{eqnarray}
0 &=&-\frac{\gamma _{c}+\gamma _{m}}{2}C_{11}-i\sqrt{6}JC_{03}-i\varepsilon
C_{10}, \\
0 &=&-\frac{3\gamma _{m}}{2}C_{03}-i\sqrt{6}JC_{11}-i\sqrt{3}\varepsilon
C_{02},
\end{eqnarray}%
\begin{eqnarray}
0 &=&-\gamma _{c}C_{20}-i2JC_{12}, \\
0 &=&-\frac{\gamma _{c}+2\gamma _{m}}{2}C_{12}-i2\sqrt{3}JC_{04}-i2JC_{20}
\notag \\
&&-i\sqrt{2}\varepsilon C_{11}, \\
0 &=&-2\gamma _{m}C_{04}-i2\sqrt{3}JC_{12}-i2\varepsilon C_{03}.
\end{eqnarray}%
To derive the optimal condition for photon blockade [i.e., $%
g_{aa}^{(2)}\left( 0\right) \approx 0$], we set $C_{20}=0$, then the
optimal effective coupling strength is obtained as%
\begin{equation}  \label{EqJ}
J_{\mathrm{opt}}=\sqrt{\frac{1}{8}\left( 2\gamma _{m}+\gamma _{c}\right)
\left( \gamma _{m}+\gamma _{c}\right) }.
\end{equation}
Substituting $\gamma_m=\gamma_c/10$ into the above equation, we have $J_{%
\mathrm{opt}} \approx 0.406 \gamma_c$, which is consistent well with the numerical
results shown in Fig.~\ref{fig2}(b).

Figure~\ref{fig4} shows the contour plot of $\log_{10}g_{aa}^{\left( 2\right)
}\left( 0\right)$ as a function of the damping rate $\gamma_m/\gamma_c$ [or $%
\gamma_c/(10\gamma_m)$] and the effective coupling strength $J/\gamma_c$.
The white dashed line in Fig.~\ref{fig4} is the optimal effective coupling
strength, given by Eq.~(\ref{EqJ}). It is clear that the optimal effective
coupling strength, given by Eq.~(\ref{EqJ}), agrees well with the numerical
results. This suggests that the strong photon blockade appearing with weak
coupling strength is induced by the destructive interference between two
different paths of two-photon excitation. We can call the interference-based
photon blockade as unconventional photon blockade, which is similar to the
unconventional photon blockade in a weakly nonlinear system of photonic
molecule~\cite%
{LiewPRL10,BambaPRA11,LemondePRA14,MajumdarPRL12,GeracePRA14,KyriienkoPRA14a,XuPRA14a,XuPRA14b,KyriienkoPRA14,ShenPRA15,ZhouPRA15}%
.

It is worth mentioning that our study is different from
that of nonlinear photonic molecules. The main difference is that
there are two separate energy scales in the weakly nonlinear photonic
molecules~\cite%
{LiewPRL10,BambaPRA11,LemondePRA14,MajumdarPRL12,KomarPRA13,GeracePRA14,KyriienkoPRA14a,XuPRA14a,XuPRA14b,KyriienkoPRA14,ShenPRA15,ZhouPRA15}%
, one is large linear coupling strength between coupled cavity modes, and the
second is small nonlinearity (up to hundred times smaller than the photon
damping rate). Here, in the quadratically coupled optomechanical systems,
there is only one parameter, i.e. the effective (nonlinear) coupling
strength $J$, which is the order of the damping rate of the optical mode
[see Eq.~(\ref{EqJ})].

\section{Correlated phonons and photons}

\begin{widetext}
\begin{figure*}[tbp]
\includegraphics[bb=32 294 565 576, width=13 cm, clip]{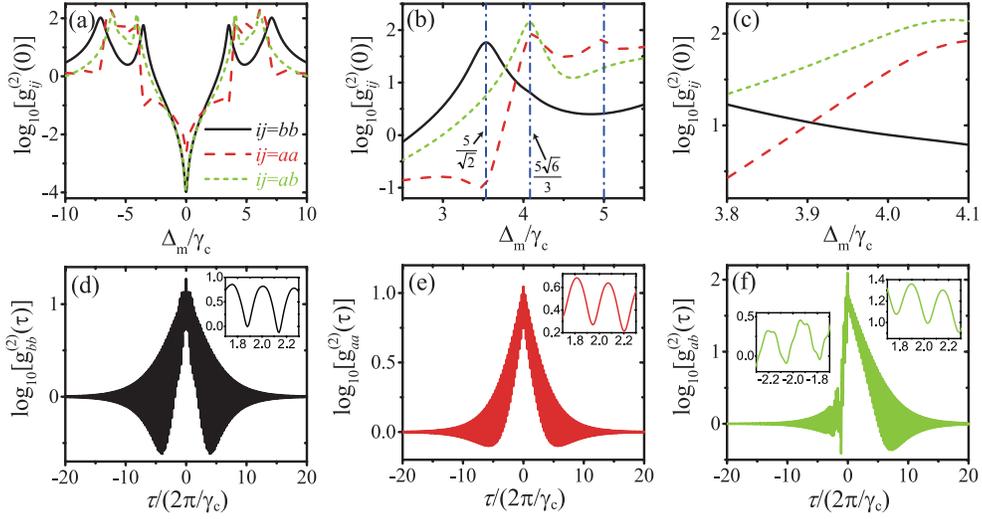}
\caption{(Color online) (a) $\log_{10}g_{ij}^{\left(2\right) }\left( 0\right)$ ($%
ij=bb,aa,ab$) is plotted as a function of the detuning $\Delta_{m}/%
\protect\gamma_c$; (b) and (c) are the local enlarged drawings of (a).
The blue dashed-dotted lines in (b) refer to $\Delta_{m}/\gamma_c=5/\sqrt{2}$, $5\sqrt{6}/3$ and $5$.
$\log_{10}g_{ij}^{\left( 2\right) }\left( \tau\right)$ [(d) $ij=bb$, (e) $ij=aa$, (f) $ij=ab$] is plotted as a function of the normalized time delay $\tau/(2\pi/\gamma_c)$ in (d)-(f) for $\Delta_{m}=3.9 \gamma_c$.
The other parameters are $\protect%
\varepsilon=0.05\protect\gamma_c$, $J = 5\protect\gamma_c$, $\omega_L=\omega_c-2\omega_0$, $\Delta=2\Delta_{m}$, $\protect\gamma_m=\protect\gamma_c/10$,
and $n_{\mathrm{th}}=10^{-4}$.}
\label{fig5}
\end{figure*}
\end{widetext}

Different from the previous two sections, here, we assume that the system
works in the strong coupling condition, i.e., $J = 5\gamma_c$. $%
\log_{10}g_{ij}^{\left(2\right) }\left( 0\right)$ ($ij=bb,aa,ab$) are
plotted as functions of the detuning $\Delta_{m}$ in Fig.~\ref{fig5}(a),
where $\omega_L=\omega_c-2\omega_0$ and $\Delta=2\Delta_{m}$ as shown in
Fig.~\ref{fig2}(c). Figures \ref{fig5}(b) and \ref{fig5}(c) are the local
enlarged drawings of Fig.~\ref{fig5}(a). As shown in Fig.~\ref{fig5}(b),
there are peaks and dips around the points $\Delta_{m}/\gamma_c=5/\sqrt{2}
$, $5\sqrt{6}/3$, and $5$ [blue dashed-dotted lines in Fig.~\ref{fig5}(b)],
which are corresponding to the resonant transitions $| 0_{0} \rangle
\rightarrow |2_{\pm 1} \rangle$, $| 0_{0} \rangle \rightarrow |3_{\pm 1}
\rangle$, and $| 0_{0} \rangle \rightarrow |4_{\pm 1} \rangle$ with energy
levels shown in Fig.~\ref{fig1}.

There is another interesting phenomenon that all the second-order
correlation functions become larger (even much larger) than $1$, i.e., $%
\log_{10}g_{ij}^{\left(2\right) }\left( 0\right)>0$ ($ij=bb,aa,ab$), in the
areas between the peaks and dips, such as the area shown in Fig.~\ref%
{fig5}(c). These imply that the photon pairs and phonon pairs can be
generated simultaneously and the generated photons and phonons are
correlated with each other. Future applications could include the two-photon
gateway, two-phonon gateway, and the correlated photon-phonon gateway~\cite%
{KubanekPRL08}.

$\log_{10}g_{ij}^{\left( 2\right) }\left( \tau\right)$ ($ij=bb,aa,ab$) are
plotted as a function of the normalized time delay $\tau/(2\pi/\gamma_c)$ in
Figs.~\ref{fig5}(d), \ref{fig5}(e) and \ref{fig5}(f) for the detuning $%
\Delta_{m}=3.9\gamma_c$. $\log_{10}g_{ij}^{\left( 2\right) }\left(
\tau\right)$ shows an oscillation behavior with the periods $%
2\pi/(n\Delta_{m})$ ($n$ is a positive integer). These oscillation behaviors
with the periods $2\pi/(n\Delta_{m})$ come from the population oscillation
between the states $| n,m \rangle \rightarrow |n,m+1 \rangle$ with detuning $%
\Delta_{m}=3.9\gamma_c$ in the weak driving condition $\varepsilon \ll
\gamma _{c}$. The time durations for the generations of phonon (photon) pairs
and correlated photons and phonons are of the order of the lifetime of the
phonons. The cross-correlation function for correlated photons and phonons
is asymmetric for $\tau>0$ and $\tau<0$.

\begin{figure}[tbp]
\includegraphics[bb=50 212 573 603, width=8.5 cm, clip]{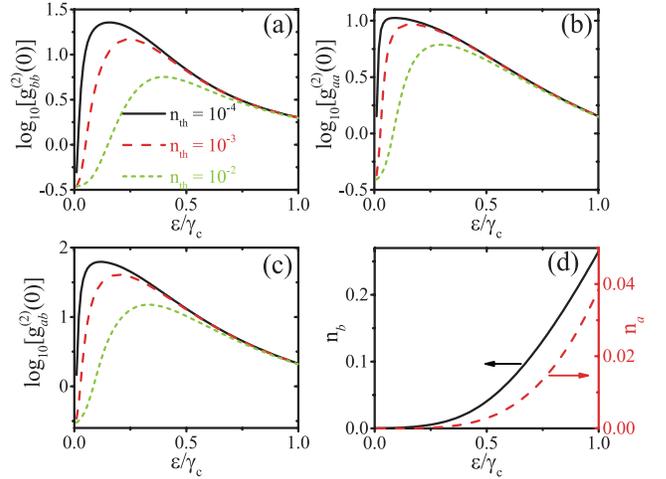}
\caption{(Color online) $\log_{10}g_{ij}^{\left( 2\right) }\left( 0\right)$
[(a) $ij=bb$, (b) $ij=aa$, (c) $ij=ab$] is plotted as a function of the
driving strength $\protect\varepsilon/\protect\gamma_c$ for different mean
thermal phonon number $n_{\mathrm{th}}$ [solid curve for $n_{\mathrm{th}%
}=10^{-4}$; dashed curve for $n_{\mathrm{th}}=10^{-3}$; dotted curve for $n_{%
\mathrm{th}}=10^{-2}$]. (d) Mean phonon number $n_b$ (solid curve) and
photon number $n_a$ (dashed curve) are plotted as functions of the driving
strength $\protect\varepsilon/\protect\gamma_c$ for mean thermal phonon
number $n_{\mathrm{th}}=10^{-4}$. The other parameters are $\Delta_{m}=3.9%
\protect\gamma_c$, $\Delta=2\Delta_{m}$, $J = 5\protect\gamma_c$, and $%
\protect\gamma_m=\protect\gamma_c/10$.}
\label{fig6}
\end{figure}

$\log_{10}g_{ij}^{\left( 2\right) }\left( 0\right)$ is plotted as a function
of the mechanical driving strength $\varepsilon/\gamma_c$ for different mean
thermal phonon numbers $n_{\mathrm{th}}$ in Fig.~\ref{fig6}: (a) $ij=bb$, (b)
$ij=aa$, (c) $ij=ab$. Mean phonon number $n_b$ and photon number $n_a$ are
plotted as functions of the mechanical driving strength $\varepsilon$ in
Fig.~\ref{fig6}(d). Similarly to the case in Fig.~\ref{fig3}, the thermal
phonons have a detrimental effect on the realization of bunching
phonons (phonons) and correlated photons and phonons. A proper increase of
the mechanical driving strength $\varepsilon$ can increase the number of
mean phonons and photons and this is also helpful to overcome the
detrimental effect induced by the thermal phonons. But if the mechanical
driving strength becomes too strong, the phonons and photons tend to behave
classically.

\section{Discussions and conclusions}

The first and also the most important condition required to observe
anticorrelated and correlated photons and phonons in quadratical coupled
optomechanical systems is the well resolved sideband limit, i.e., $\omega_m
\gg \gamma_c$. This requirement could be reached for the optomechanical
crystals as indicated in Ref.~\cite{ParaisoPRX15} by numerical simulations, where
the simulated parameters are: mechanical resonance frequency $%
\omega_m/2\pi=225$ MHz and optical damping rate $\gamma_c/2\pi=20$ MHz.
Another candidate system is the Fabry-Perot cavity with
membrane-in-the-middle~\cite%
{ThompsonNat08,Flowers-JacobsAPL12,HXuNat16,HXuNC17} and the resolved
sideband limit was reached in Ref.~\cite{HXuNat16} with mechanical resonance
frequency $\omega_m/2\pi\approx 788$ kHz and optical damping rate $%
\gamma_c/2\pi=177$ kHz.
In addition, the quadratical couplings have been explored in a number of
other optomechanical systems~\cite%
{PurdyPRL10,JTHill13,DoolinPRA14,BrawleyNC16}.
Secondly, it is a outstanding challenge to detect single phonons directly in the experiments. The measurements of the correlation of the phonons and the cross-correlation between photons and phonons can be realized by converting the mechanical signals into optical signals through auxiliary optomechanical couplings~\cite{RamosPRL13,CohenNature15,XWXuPRA16}, which have been realized in a recent experiment~\cite{CohenNature15}.

In summary, we have studied the photon, phonon statistics and the
cross-correlation between photons and phonons in a quadratically coupled
optomechanical system. We show that photon blockade, phonon blockade, and
strong anticorrelation between photons and phonons can be observed in the same
parameter area. Phonon blockade and strong anticorrelation between photons and
phonons can be understood by the nonlinear energy spectrum of the system,
while the photon blockade with weak nonlinear coupling strength can only be
explanted by the destructive interference between different paths for
two-photon excitation. The combination of photon blockade, phonon blockade,
and strongly anticorrelated photons and phonons provides us a way to
generate anticorrelated single photons and single phonons. Further more, in
the strongly nonlinear coupling condition, photon pairs and phonon pairs can
be generated simultaneously, and the photon and phonon pairs are correlated
with each other, which can be used to generate two-photon gateway,
two-phonon gateway, and the correlated photon-phonon gateway~\cite%
{KubanekPRL08}.

\vskip 2pc \leftline{\bf Acknowledgement}

X.W.X. is supported by the National Natural Science Foundation of China
(NSFC) under Grants No.11604096 and the Startup Foundation for Doctors of
East China Jiaotong University under Grant No. 26541059. A.X.C. is supported
by NSFC under Grant No. 11775190. Y.X.L. is supported by the National Basic
Research Program of China(973 Program) under Grant No. 2014CB921401, the
Tsinghua University Initiative Scientific Research Program, and the Tsinghua
National Laboratory for Information Science and Technology (TNList)
Cross-discipline Foundation.

\bibliographystyle{apsrev}
\bibliography{ref}

\end{document}